# Automated lung segmentation from CT images of normal and COVID-19 pneumonia patients

Faeze Gholamiankhah[1+], Samaneh Mostafapour[2+], Nouraddin Abdi Goushbolagh[1], Seyedjafar Shojaerazavi[3], Parvaneh Layegh[4], Seyyed Mohammad Tabatabaei[5, 6], Hossein Arabi[7]

[1]Department of Medical Physics, Faculty of Medicine, Shahid Sadoughi University of Medical Sciences, Yazd, Iran

[2] Department of Radiology Technology, Faculty of Paramedical Sciences, Mashhad University of Medical Sciences, Mashhad, Iran

[3] Department of Cardiology, Ghaem Hospital, Mashhad University of Medical Sciences, Mashhad, Iran

[4] Department of Radiology, Faculty of Medicine, Mashhad University of Medical Sciences, Mashhad, Iran

[5] Department of Medical Informatics, School of Medicine, Mashhad University of Medical Sciences, Mashhad, Iran

[6] Clinical Research Development Unit, Imam Reza Hospital, Mashhad University of Medical Sciences, Mashhad, Iran

[7]Division of Nuclear Medicine and Molecular Imaging, Geneva University Hospital, CH-1211 Geneva 4, Switzerland

**Abstract**

**Objective:**

Automated semantic image segmentation is an essential step in quantitative image analysis and disease diagnosis. This study investigates the performance of a deep learning-based model for lung segmentation from CT images for normal and COVID-19 patients.

**Methods:**

Chest CT images and corresponding lung masks of 1200 confirmed COVID-19 cases were used for training a residual neural network. The reference lung masks were generated through semi-automated/manual segmentation of the CT images. The performance of the model was evaluated on two distinct external test datasets including 120 normal and COVID-19 subjects, and the results of these groups were compared to each other. Different evaluation metrics such as dice coefficient (DSC), mean absolute error (MAE), relative mean HU difference, and relative volume difference were calculated to assess the accuracy of the predicted lung masks.

**Results:**

The proposed deep learning method achieved DSC of 0.980 and 0.971 for normal and COVID-19 subjects, respectively, demonstrating significant overlap between predicted and reference lung masks. Moreover, MAEs of 0.037 HU and 0.061 HU, relative mean HU difference of -2.679% and -4.403%, and relative volume difference of 2.405% and 5.928% were obtained for normal and COVID-19 subjects, respectively. The comparable performance in lung segmentation of the normal and COVID-19 patients indicates the accuracy of the model for the identification of the lung tissue in the presence of the COVID-19 induced infections (though slightly better performance was observed for normal patients).

**Conclusion:**

The promising results achieved by the proposed deep learning-based model demonstrated its reliability in COVID-19 lung segmentation. This prerequisite step would lead to a more efficient and robust pneumonia lesion analysis.

# Introduction

The novel coronavirus named SARS-CoV-2 was first broke out in Wuhan China in December 2019 and has invaded most countries around the globe [1, 2]. Through infecting the respiratory tracts, this virus causes respiratory syndromes [3], where the real-time polymerase chain reaction (RT-PCR) is known as the common/standard method for the diagnosis of COVID-19. However, some concerns such as a high false-negative rate when the viral load is low in the test specimen have limited its application [1, 4].

On the contrary, chest X-ray or computed tomography (CT) imaging are considered as faster and complementary procedures that facilitate the early screening of COVID-19 infections [5, 6]. However, CT imaging outperforms X-ray radiography in providing more structural/anatomical details of the lung [7-13]. Chest CT images are reported to have a sensitivity of 0.97 in the diagnosis of COVID-19 [14], enabling to detect the radiological patterns like bilateral and peripheral ground-glass opacities and patchy consolidations in the lung of infected patients [2, 15]. Moreover, quantitative analysis of CT images provides key information about the size of lesions and severity of the disease [16] wherein reliable lung CT image segmentation is a critical prerequisite step in this regard [17, 18].

Different approaches for lung segmentation have been adopted including manual segmentation, rule-based, atlas-based, machine learning-based as well as hybrid techniques [19-21]. Manual segmentation is a time-consuming and labor-intensive task particularly in situations that the health system is overloaded [6]. Other conventional methods such as atlas-based or intensity-based algorithms lead to acceptable results in normal cases or mild disease, however, their implementation in diseases like COVID-19 that infection alters the common pattern/structure of the lung is inefficient [19, 22, 23]. To address this challenge, recent researches have evaluated the use of deep learning models for lung and lesion segmentation and demonstrated the promising performance of convolutional neural networks in distinguishing lung from the chest wall [24, 25].

There are a number of studies that have employed the common deep learning-based image segmentation architectures such as U-Net, 3D U-Net, U-Net++, and V-Net for COVID-19 lung segmentation [6, 26, 27]. Furthermore, some studies developed and evaluated state-of-the-are algorithms for the cases with insufficient annotated datasets using transfer learning or weakly annotated datasets [28-30]. Due to the presence of considerable abnormalities in the lung caused by the COVID-19 infection, segmentation of the lung in COVID-19 patients faces the challenge of lung boundary and infection discrimination compared to the normal patients which bear distinct contrast between the lung tissue and chest wall [31].

This study sought to assess the efficiency of the deep learning approach in automated lung segmentation from CT images of patients with COVID-19 in comparison with normal patients. Automated lung segmentation would assist quantitative analysis and segmentation of infections by removing unnecessary regions in the chest images. To conduct a meticulous investigation, a dataset of normal patients was also employed in addition to CT scans of infected patients to study the impact of lung abnormalities caused by COVID-19 infection on the lung segmentation.

**Material and Methods**

**Dataset**

The dataset used in this study consists of chest CT images from 1200 patients with RT-PCR confirmed COVID-19 and 120 normal patients without any lung abnormalities. CT image acquisition was performed on a Siemens Somatom Spirit Dual Slice CT scanner with tube energy of 130 kVp, tube current of 48 mAs, rotation time (TI) of 0.8 s, and slice thickness of 5 mm. For generating ground truth lung masks, CT images were segmented semi-automatically using Pulmonary Toolkit (PTK) software [32] and the resulting binary masks were manually corrected to avoid any noticeable errors. Prior to the training of the network, all images were cropped to eliminate areas outside of the lung volume and resized to a matrix size of 296×216 voxels by a linear interpolation algorithm. Thereafter, Hounsfield units were scaled to an intensity range between 0 and 1.

**Implementation detail**

The ResNet model implemented in NiftyNet was utilized for the implementation of the automated lung segmentation. NiftyNet is an open-source platform built upon TensorFlow that consists of common convolutional neural networks used in medical imaging [33]. The ResNet architecture, as shown in figure 1, comprises 20 convolutional layers wherein every two layers are connected together by residual connections. In this network, dilation factors of one, two, and four are applied on the convolutional kernels to extract low-level, mid-level, and high-level features from the input images, respectively. Also, a fully connected softmax layer is embedded as the last layer of the network [34, 35].

From the total number of subjects included in this study, 1080 CT images and their corresponding masks were randomly selected for the training of the network, and the remaining 120 subjects for validation (external validation). To make sure that there is no risk of overfitting, 5% of the training subjects were employed for validation of the model within the training phase. The investigations revealed no considerable difference between training and evaluation losses.

The training of the deep learning model was performed on two-dimensional slices using the following settings: learning rate = 0.02, optimizer = Adam, loss function = Dice_NS, decay = 0.0001, batch size = 17, and weights regression type = L2norm.

**Evaluation Metrics**

To evaluate the performance of the deep learning model, predicted and ground truth lung segmentations were compared on the external test dataset including 120 COVID-19 patients and 120 normal subjects. The assessment was performed via calculating the dice similarity coefficient (DSC) (Eq. 1), Jaccard index (JC) (Eq. 2), mean error (ME) (Eq. 3), mean absolute error (MAE) (Eq. 4) within the estimated lung region.

$$DSC = \frac{2 \mid I_r \cap I_p \mid}{\mid I_r \mid + \mid I_p \mid} \quad (1)$$

$$JC = \frac{\mid I_r \cap I_p \mid}{\mid I_r \cup I_p \mid} \quad (2)$$

$$ME = \frac{1}{N} \sum_{i=1}^{N} (I_p(i) - I_r(i)) \quad (3)$$

$$MAE = \frac{1}{V} \sum_{i=1}^{V} \mid I_p(i) - I_r(i) \mid \quad (4)$$

Here, $I_r$ and $I_p$ denote the reference and predicted lung masks. $V$ and $i$ indicate the total number of voxels in the lung area and the index of voxels in $I_r$ and $I_p$ images, respectively.

Moreover, the false-positive ratio (Eq. 5) and false-negative ratio (Eq. 6) were estimated as follow:

$$False\ Positive\ Ratio = \frac{FP}{FP + TN} \quad (5)$$

$$False\ Negative\ Ratio = \frac{FN}{FN + TP} \quad (6)$$

In Eqs. 5 and 6, *FP* is the number of false positives, *TN* is the number of true negatives, *FN* is the number of false negatives and *TP* is the number of true positives for the voxels residing in the lung area.

Furthermore, relative mean CT number (Hounsfield Unit (HU)) difference, absolute relative mean HU difference, relative volume difference, and absolute relative volume difference metrics were calculated between the reference and predicted lung volumes.

## Results

Representative results of the lung segmentation for normal and COVID-19 subjects are presented in figures 2 and 3. Figure 2 shows a good match between the reference and predicted masks for both normal and infected lung tissues, which indicates the promising performance of the deep learning model in lung border detection. Figure 3 depicts minor segmentation errors for two cases in which the model was not successful to define an accurate margin for the lung and excluding bronchus from the segmented area (outlier report). The miss-segmentation error is more noticeable in COVID-19 patients due to the similar intensity of the chest wall and severe infections, which have rendered the accurate identification of the lung boundary very challenging.

Table 1 summarizes the minimum, maximum, mean, and standard deviation of the quantitative metrics for lung segmentation including DSC, JC, ME, MAE, False Positive Ratio, False Negative ratio, mean HU difference within the lung mask, and volume difference calculated for external validation datasets. Overall, the proposed model showed better performance in lung segmentation of normal subjects compared to COVID-19 patients due to the high-density infections residing close to the chest wall in COVID-19 subjects.

The mean DSC and JC values in normal group were 0.980 ± 0.003 and 0.962 ± 0.007, and in COVID-19 group were 0.971 ± 0.017 and 0.938 ± 0.040, respectively. The deep learning model achieved ME of -0.015 ± 0.009 HU and -0.024±0.042 HU, and MAE of 0.037±0.007 HU and 0.061±0.040 HU within the lung mask for the normal and COVID-19 subjects, respectively. Moreover, comparable measures of false positive and false negative ratios were obtained for the two test datasets. The quantitative assessment revealed relative Mean HU Differences of (-2.679±0.382% and -4.403±4.097%) and relative volume differences of (2.405±7.359% and 5.928±17.261%) calculated for the lung tissue in the normal and COVID-19 subjects, respectively.

Figure 4 illustrates boxplots of DSC, JC, ME, MAE, Relative Mean HU Difference, and Relative Volume Difference metrics calculated for the two test datasets. Overall, considering the entire metrics, larger standard deviation and outliers were observed in the COVID-19 patients compared to the normal subjects.

## Discussion

Recent studies suggest that chest CT imaging findings play a significant role in COVID-19 diagnosis and management [36, 37]. Accurate lung segmentation is a crucial step for the calculation of the quantitative indices, measurement of lung engagement, and disease severity [38, 39]. The existing segmentation methods, that have shown satisfactory performance in normal or mild lung diseases, are either time-consuming or face serious challenges in the segmentation of COVID-19 infected lung tissue due to the close similarity between infections and normal tissues [7]. Deep learning-based models have been widely adopted lately by researchers as a dependable solution to assist clinicians in fast and efficient COVID-19 lung and lesion segmentation [40, 41]. This study set out to investigate the performance of a state-of-the-art deep learning approach in lung segmentation. The network was evaluated on two external test datasets including normal and COVID-19 subjects. Multiple evaluation metrics were used to perform a comprehensive performance assessment of the model.

Gerard et al. [22] developed a segmentation algorithm called LobeNet aiming at the prediction of the left and right lung regions in the presence of diffuse opacities and consolidations. The quantitative assessment of LobeNet performance on 87 patients with COVID-19 showed an average Dice coefficient of 0.985. Moreover, Ma et al. [38] reported a DSC score of 0.973 and 0.977 obtained from a 3D U-Net for the left and right lung segmentation from COVID-19 CT images. Tilborghs et al. [40] compared the performance of various deep learning algorithms using a multicenter COVID-19 dataset. They concluded that combining different methods could improve the segmentation performance and increased the Dice coefficient up to 0.987. The proposed ResNet model implemented in this study achieved average Dice coefficients of 0.980±0.003 and 0.971±0.017 in normal and COVID-19 external test data set, respectively. This indicates a very good segmentation overlap between predicted and ground truth lung masks.

Yan et al. [6] established a new network called COVID-SegNet for the segmentation of CT images with COVID-19 infection. They compared their proposed method against other state-of-the-art models including FCN, UNet, VNet, and UNet++. The COVID-SegNet achieved DSC, sensitivity, and precision of 0.865, 0.986, and 0.983, respectively. In another study conducted by Trivizakis et al. [42], the U-Net architecture, one of the most common image segmentation architectures, was implemented for COVID-19 lung segmentation. They reported a DSC of 0.950, sensitivity of 0.920, and specificity of 0.875. Muller et al. [43] performed a similar analysis using 3D U-Net for lung segmentation using a cross-validation scheme to reduce/avoid the risk of overfitting. Their approach led to lung segmentation with DSC of 0.956, sensitivity of 0.956, and specificity of 0.998. Compared to the results reported in the literature, the false-positive ratio of 0.005 and 0.007, and false-negative ratio of 0.026 and 0.044 obtained for the normal and COVID-19 patients in this study demonstrated the promising performance of the proposed deep learning approach.

The investigation of the deep learning model showed high accuracy in terms of Relative Mean HU Difference (-2.679% and -4.403%), and Relative Volume Difference (2.405% and 5.928%) in normal and COVID-19 cases, respectively. Therefore, the performance of the ResNet model in COVID-19 patients, wherein the infection has changed the intensity of the lung, is comparable with its performance in normal subjects.

The investigation of the framework proposed in this work revealed promising results in lung segmentation of patients with COVID-19. However, there are some limitations associated with this study that require to be addressed in future studies. One of these limitations would be the lack of a multicenter dataset to investigate the sensitivity of the developed model to the variation of the image quality and acquisition parameters across different scanners/centers. Moreover, in order to develop a comprehensive framework which is suitable for clinical practice, the performance of the ResNet model in lesion segmentation should be evaluated. This would enable distinguishing COVID-19 lesions from other pneumonia, and diagnosing subtypes of COVID-19 pneumonia.

**Conclusion**

In this work, deep learning-based automated lung segmentation from chest CT images in patients with COVID-19 was investigated. The proposed method achieved very promising results with a dice coefficient of 0.980 and 0.971 in normal and COVID-19 external test datasets, respectively. The reliable lung segmentation would facilitate lesion segmentation and will lead to more accurate quantitative analysis, diagnosis, and treatment of the COVID-19 patients.

**Table1.** Statistics of quantitative metrics including ME, MAE, RMSE, RE, RVD, Dice, JC, Sensitivity, SSIM, and PSNR calculated between the reference and predicted lung masks in normal and COVID-19 test datasets.

| Parameter | Normal | | | COVID-19 | | |
|---|---|---|---|---|---|---|
| | Min | Max | Mean ± SD | Min | Max | Mean ± SD |
| Dice Coefficient | 0.970 | 0.985 | 0.980 ± 0.003 | 0.903 | 0.986 | 0.971 ± 0.017 |
| Jaccard Index | 0.942 | 0.971 | 0.962 ± 0.007 | 0.823 | 0.973 | 0.938 ± 0.040 |
| ME | -0.031 | 0.015 | -0.015 ± 0.009 | -0.155 | 0.042 | -0.024 ± 0.042 |
| MAE | 0.028 | 0.057 | 0.037 ± 0.007 | 0.026 | 0.176 | 0.061 ± 0.040 |
| False Positive Ratio | 0.001 | 0.022 | 0.005 ± 0.004 | 0.003 | 0.028 | 0.007 ± 0.004 |
| False Negative Ratio | 0.020 | 0.036 | 0.026 ± 0.003 | 0.011 | 0.167 | 0.044 ± 0.040 |
| Relative Mean HU Diff (%) | -3.673 | -2.020 | -2.679 ± 0.382 | -16.799 | -1.183 | -4.403 ± 4.097 |
| Absolute Relative Mean HU Diff (%) | 0 | 0.374 | 0.828 ± 1.318 | 0.921 | 10.894 | 3.253 ± 3.106 |
| Relative Volume Diff (%) | -4.726 | 29.524 | 2.405 ± 7.359 | -12.666 | 90.561 | 5.928 ± 17.261 |
| Absolute Relative Volume Diff (%) | 1.960 | 33.164 | 7.875 ± 6.548 | 2.199 | 91.486 | 12.743 ± 16.384 |

**Figure 1.** The architecture of the ResNet model.

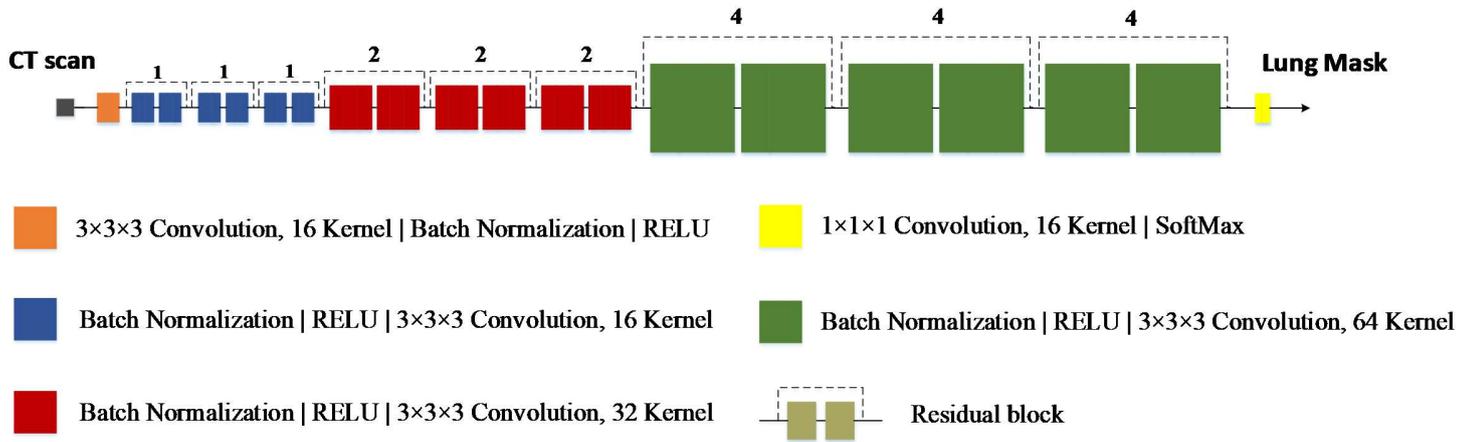

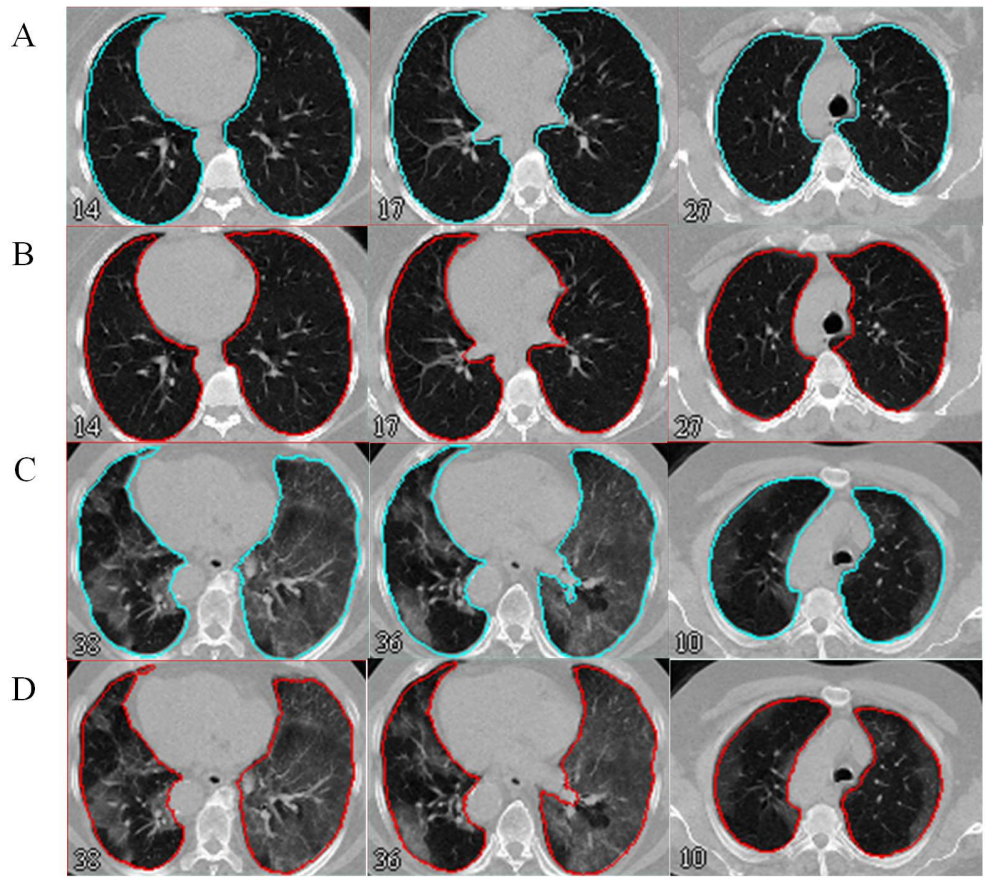

**Figure 2.** Axial views of A) Ground truth lung masks in normal subjects. B) and the corresponding predicted lung masks. C) Ground truth lung masks in COVID-19 subjects D) and the corresponding predicted lung masks.

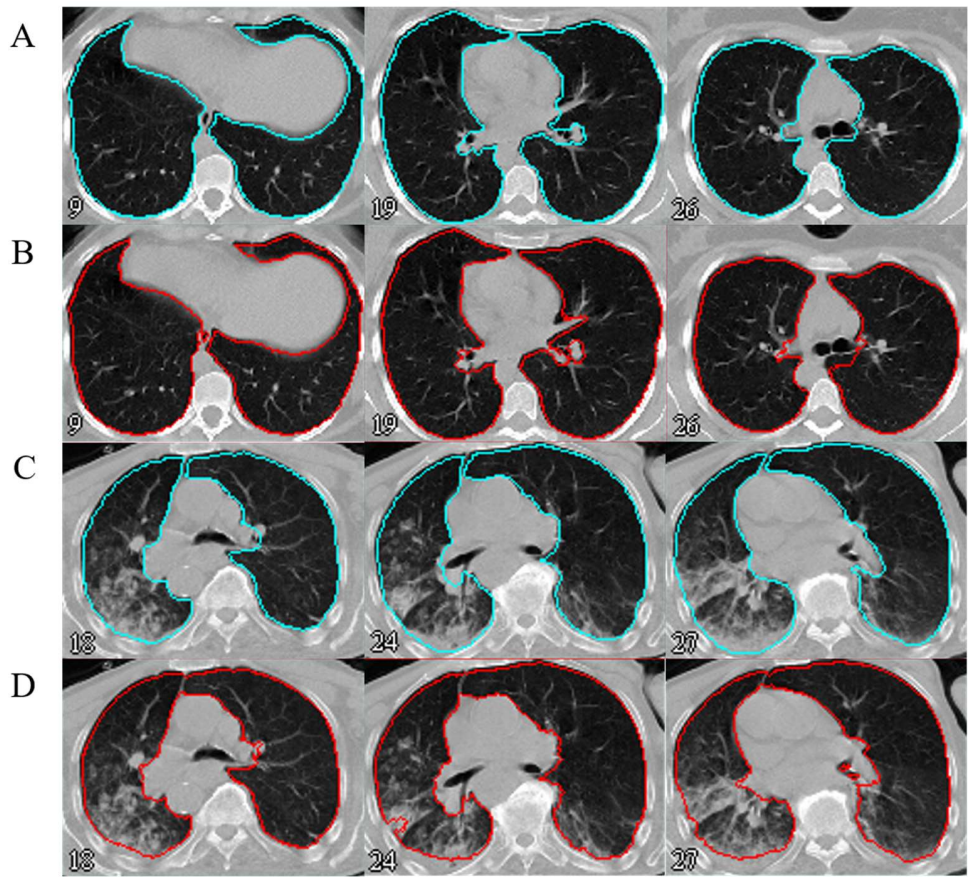

**Figure 3.** Outlier report. Cases with noticeable errors. Axial views of A) Ground truth lung masks in normal subjects. B) and the corresponding predicted lung masks. C) Ground truth lung masks in COVID-19 subjects D) and the corresponding predicted lung masks.

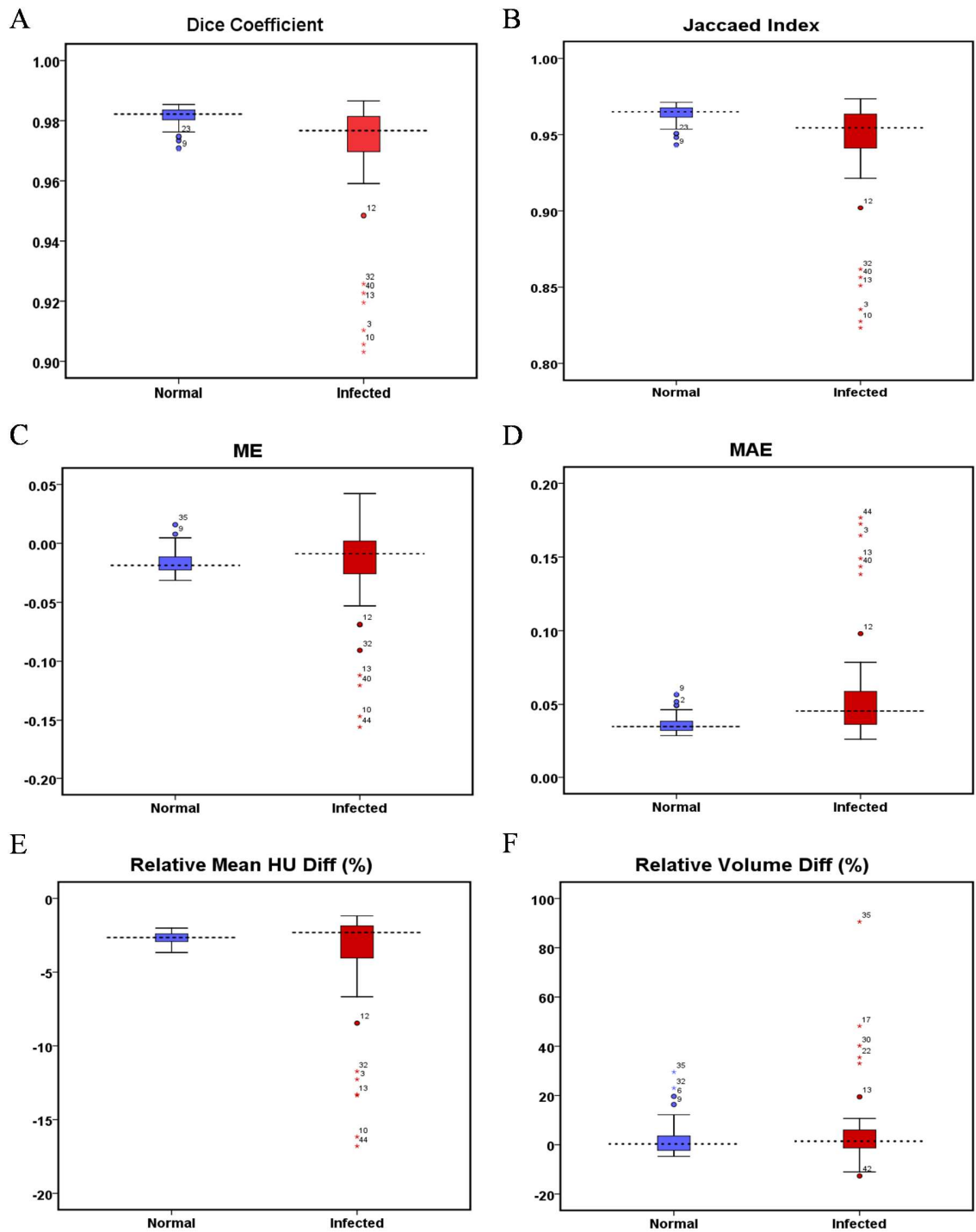

**Figure 4.** Box plots comparing A) Dice Coefficient, B) Jaccard Index, C) ME, D) MAE, E) Relative Mean HU Difference (%) and F) Relative Volume Difference(%) metrics between normal and COVID-19 test datasets.